\documentstyle[sprocl,epsfig]{article}
\def\Journal#1#2#3#4{{#1} {\bf #2}, #3 (#4)}

\def\NPB{{\em Nucl. Phys.} B}
\def\PLB{{\em Phys. Lett.}  B}
\def\APP{{\em Acta Phys. Pol.}  B}
\def\PRL{\em Phys. Rev. Lett.}

\def\ZPC{{\em Z. Phys.} C}
\def\EPJ{{\em E. Phys. J.} C}

\def\be{\begin{equation}}
\def\ee{\end{equation}}
\def\bea{\begin{eqnarray}}
\def\eea{\end{eqnarray}}
\def\Z0{$\mathrm{Z}$}
\def\MZ{$\mathrm{M}_{\mathrm{Z}}$}
\def\MW{$\mathrm{M}_{\mathrm{W}}$}
\def\MH{$\mathrm{M}_{\mathrm{H}}$}
\def\VJK{$\mathrm{V}_{\mathrm{jk}}$}
\def\mt{$\mathrm{m}_{\mathrm{t}}$}
\def\GM{$G_{\mu}$}
\def\als{$\alpha_{\mathrm{s}}$}
\def\SEF{$\sin^2\theta_{\mathrm{eff}}$}
\def\GZ{$\Gamma_{\mathrm{Z}}$}
\def\Gl{$\Gamma_{\mathrm{l}}$}
\def\Gh{$\Gamma_{\mathrm{h}}$}
\def\Rb{$\mathrm{R}_{\mathrm{b}}$}
\def\Rc{$\mathrm{R}_{\mathrm{c}}$}
\def\Rl{$\mathrm{R}_{\mathrm{l}}$}
\def\S0{$\sigma^0_{\mathrm{f\bar{f}}}$}
\def\ALR{$\mathrm{A}_{\mathrm{LR}}$}
\begin{document}

%==============================================================================
\flushbottom
\begin{titlepage}
%==============================================================================
%\begin{center}
%\large {EUROPEAN ORGANIZATION FOR PARTICLE PHYSICS\\}
%\end{center}
%\vspace*{0.2cm}
%\begin{flushright}
%       CERN-OPEN/98-xxx\\
%       xx November 1998\\
%\end{flushright}

\vspace*{1cm}
\begin{center}
\boldmath
\Huge {\bf Precision test of the Standard Model from \\
               Z physics \\[.5cm]
}
\vspace*{0.5cm}
\large\textbf{
Frederic Teubert \\}
\vspace*{0.5cm}
\large{
European Laboratory for Particle Physics (CERN), \\
CH-1211 Geneva 23, Switzerland \\
E-mail: frederic.teubert@cern.ch \\
}
\unboldmath
\end{center}
\vspace*{0.5cm}
\begin{abstract}
The measurements performed at LEP and SLC 
have substantially improved the precision of the test of the 
Minimal Standard Model. The precision is such that there 
is sensitivity to pure weak radiative corrections. This allows
to indirectly determine the top mass (\mt=161$\pm$8~GeV),  
the W-boson mass (\MW=80.37$\pm$0.03~GeV), and to set an upper limit on the
the Higgs boson mass of 262~GeV at 95\% confidence level.
\end{abstract}
\vspace*{1.5cm}
\begin{center}
\large{\em
Invited talk presented at the IVth International Symposium \\
on Radiative Corrections (RADCOR 98), \\
Barcelona, Spain, Catalonia, September 8-12, 1998.}\\
\end{center}

\end{titlepage}
\setcounter{page}{2}
\renewcommand{\thefootnote}{\arabic{footnote}}
\setcounter{footnote}{0}

\title{PRECISION TESTS OF THE STANDARD MODEL FROM \Z0~PHYSICS}

\author{F.TEUBERT}

\address{European Laboratory for Particle Physics (CERN), 
\\ CH-1211 Geneva 23, Switzerland\\E-mail: frederic.teubert@cern.ch}


\maketitle\abstracts{ The measurements performed at LEP and SLC 
have substantially improved the precision of the test of the 
Minimal Standard Model. The precision is such that there 
is sensitivity to pure weak radiative corrections. This allows
to indirectly determine the top mass (\mt=161$\pm$8~GeV),  
the W-boson mass (\MW=80.37$\pm$0.03~GeV), and to set an upper limit on the
the Higgs boson mass of 262~GeV at 95\% confidence level. }

\section{\bf Introduction}

In the context of the Minimal Standard Model (MSM), any ElectroWeak (EW) process 
can be computed at tree level from $\alpha$ (the fine structure constant 
measured at values of $q^2$ close to zero), \MW~(the W-boson mass), 
\MZ~(the Z-boson mass), and \VJK~(the 
Cabbibo-Kobayashi-Maskawa flavour-mixing matrix elements).  

When higher order corrections are included, any observable can be predicted in the 
``on-shell'' renormalization scheme as a function of:

\noindent
\bea
O_i & = & f_i(\alpha, \alpha_{\mathrm{s}},\mathrm{M}_{\mathrm{W}} ,\mathrm{M}_{\mathrm{Z}},
\mathrm{M}_{\mathrm{H}},\mathrm{m}_{\mathrm{f}},\rm{V}_{\rm{jk}})  \nonumber 
\eea

\noindent and contrary to what happens with ``exact gauge symmetry theories'', 
like QED or QCD, the effects of heavy particles do not decouple. Therefore, the 
MSM predictions depend on the top mass 
%($\frac{\mathrm{m}^2_{\mathrm{t}}}{\mathrm{M}^2_{\mathrm{Z}}}$) 
($\mathrm{m}^2_{\mathrm{t}}/\mathrm{M}^2_{\mathrm{Z}}$) 
and to less extend to the Higgs mass 
%(log($\frac{\mathrm{M}^2_{\mathrm{H}}}{\mathrm{M}^2_{\mathrm{Z}}}$)),
(log($\mathrm{M}^2_{\mathrm{H}}/\mathrm{M}^2_{\mathrm{Z}}$)),
or to any kind of ``heavy new physics''. 

The subject of this letter is to show how the high precision achieved in the 
EW measurements from \Z0~physics allows to test the MSM beyond the tree level 
predictions and, therefore, how this measurements are able to indirectly determine 
the value of \mt~and \MW, 
to constrain the unknown value of \MH, and at the same time to 
test the consistency between measurements and theory. At present the uncertainties 
in the theoretical predictions are dominated by the precision on the input 
parameters.

\subsection{\bf Input Parameters of the MSM}\label{subsec:input_par}

The W mass is one of the input parameters in the ``on-shell'' renormalization scheme.
It is known with a precision of about 0.07\%, although the usual procedure is 
to take \GM~(the Fermi constant measured in the muon decay) 
to predict \MW~as a function of the rest of the input parameters 
and use this more precise value. 

Therefore, the input parameters are chosen to be: 

\noindent
\bea
 \rm{Input~Parameter}       &    \rm{Value}              & \rm{Relative~Uncertainty}       \nonumber \\
%\alpha^{-1}(0)            = & 137.0359895 (61)                             &  10^{-6} \%  \nonumber \\ 
\alpha^{-1}(\mathrm{M}^2_{\mathrm{Z}})  = & 128.896 (90)                    &  0.07    \%  \nonumber \\ 
\alpha_{\mathrm{s}}(\mathrm{M}^2_{\mathrm{Z}}) = & 0.119 (2)                &  1.7     \%  \nonumber \\ 
 G_{\mu}                  = & 1.16639 (2) \times 10^{-5}~\mathrm{GeV^{-2}}  &  0.0017  \%  \nonumber \\ 
\mathrm{M}_{\mathrm{Z}}   = & 91.1867 (21)~\mathrm{GeV}                     &  0.0023  \%  \nonumber \\
\mathrm{m}_{\mathrm{t}}   = & 173.8   (50)~\mathrm{GeV}                     &  2.9     \%  \nonumber \\
\mathrm{M}_{\mathrm{H}}   > & 89.8~\mathrm{GeV}~@95\%~C.L.                  &              \nonumber 
\eea

Notice that the less well known parameters are \mt, \als~and, of course, 
the unknown value of \MH. The next less well known parameter is 
$\alpha^{-1}(\mathrm{M}^2_{\mathrm{Z}})$, even though its value at 
$q^2 \sim 0$ is known with an amazing relative precision of $4 \times 10^{-8}$,
($\alpha^{-1}(0)=137.0359895~(61)$). 

The reason for this loss of precision when one computes the running of $\alpha$,

\noindent
\bea
 \alpha^{-1}(\mathrm{M}^2_{\mathrm{Z}}) & = & \frac{\alpha^{-1}(0)}{1-\Pi_{\gamma\gamma}}  \nonumber
\eea

\noindent is the large contribution from the light fermion loops to the photon 
vacuum polarisation, $\Pi_{\gamma\gamma}$. 
The contribution from leptons and top quark loops is well calculated  
but for the light quarks non-perturbative
QCD corrections are large at low energy scales. The method so far
has been to use the measurement of the hadronic cross section 
through one-photon exchange, normalised to the point-like muon 
cross-section, R(s), and evaluate the dispersion integral:

\noindent
\bea
\Re(\Pi^{\mathrm{had}}_{\gamma\gamma}) & = & \frac{\alpha \mathrm{M}^2_{\mathrm{Z}}}{3 \pi} \Re 
\left( \int \frac{R(s')}{s'(s'-\mathrm{M}^2_{\mathrm{Z}}+i\epsilon)} ds' \right) 
\eea

\noindent giving~\cite{OLDALPHA} $\Pi_{\gamma\gamma} = - 0.0632 \pm 0.0007$, the error being 
dominated by the experimental uncertainty in the cross section measurements.

Recently, several new {\em ``theory driven''} calculations~\cite{DAVIER}~\cite{NEWALPHA} 
have reduced this error by a factor of about 4.5,
% ($0.016 \%$~relative precision),
by extending the regime of applicability of Perturbative QCD (PQCD). This needs to 
be confirmed using precision measurements of the hadronic cross section at 
$\sqrt{s} \sim$~1~to~7~GeV. The very preliminary first results from 
BESS II~\cite{BESS} seem to validate this procedure, being in agreement with 
the predictions from PQCD.
 
\subsection{\bf What are we measuring to test the MSM?}\label{subsec:what_meas}

From the point of view of radiative corrections we can divide the \Z0~measurements 
into three different groups: the \Z0~total and partial widths,
% (\GZ,\Gl,\Gh,...), 
the partial width into b-quarks ($\Gamma_{\mathrm{b}}$), and the \Z0~asymmetries (\SEF). 
For instance,
the leptonic width (\Gl) is mostly sensitive to isospin-breaking loop corrections 
($\Delta \rho$), the asymmetries are specially sensitive to radiative corrections to 
the \Z0~self-energy, and \Rb~is mostly sensitive to vertex corrections in the decay 
$\mathrm{Z} \rightarrow b \bar{b}$. One more parameter,
$\Delta r$, is necessary to describe the radiative corrections to the relation 
between \GM~and \MW.

The sensitivity of these three \Z0~observables and \MW~to the input parameters is shown in 
table~\ref{tab:sensitiv}. The most sensitive observable to the unknown 
value of \MH~are the \Z0~asymmetries parametrised via \SEF. However also the sensitivity 
of the rest of the observables is very relevant compared to the achieved experimental precision.

\begin{table}[t]
\caption{ Relative error in units of per-mil on the MSM predictions 
induced by the uncertainties on the input parameters. The second column shows 
the present experimental errors.\label{tab:sensitiv}}
\vspace{0.2cm}
\begin{center}
\footnotesize
\begin{tabular}{|l|c|c c c c|}
\hline
{} &\raisebox{0pt}[13pt][7pt]{Exp. error} &
\raisebox{0pt}[13pt][7pt]{$\Delta$\mt = 5.0~GeV} & \raisebox{0pt}[13pt][7pt]{$\Delta$\MH [90-1000]~GeV} &
\raisebox{0pt}[13pt][7pt]{$\Delta$\als = 0.002} & \raisebox{0pt}[13pt][7pt]{$\Delta\alpha^{-1}$ = 0.090} \\
\hline
$\Gamma_{\mathrm{Z}}$      &  1.0  &      0.5   & \bf{3.4}  &  0.5  &  0.3 \\
$\mathrm{R}_{\mathrm{b}}$  &  3.4  &  \bf{0.8}  &     0.1   &   -   &   -  \\
\MW                        &  0.7  &      0.4   & \bf{2.2}  &   -   &  0.2 \\
\SEF                       &  0.8  &      0.7   & \bf{5.8}  &   -   &  1.0 \\
\hline
\end{tabular}
\end{center}
\end{table}

\section{\bf \Z0~lineshape}

The shape of the resonance is completely characterised by three parameters: the position of the 
peak (\MZ), the width (\GZ) and the height (\S0) of the resonance:

\noindent
\bea
\sigma^0_{\mathrm{f\bar{f}}} & = & \frac{12\pi}{\mathrm{M}^2_{\mathrm{Z}}} 
         \frac{\Gamma_{\mathrm_{e}}\Gamma_{\mathrm_{f}}}{\Gamma^2_{\mathrm_{Z}}} 
\eea

The good capabilities 
of the LEP detectors to identify the lepton flavours allow to measure the ratio of 
the different lepton species with respect to the hadronic cross-section, 
\Rl = $\frac{\Gamma_{\mathrm{h}}}{\Gamma_{\mathrm{l}}}$. 

About 16~million \Z0~decays have been analysed by the four LEP collaborations, 
leading to a statistical precision on \S0 of 0.03 \% ! Therefore, 
the statistical error is not the limiting factor, but more the experimental systematic 
and theoretical uncertainties.

The error on the measurement of \MZ~is dominated by the uncertainty on the absolute 
scale of the LEP energy measurement (about 1.7~MeV), 
while in the case of \GZ~it is the point-to-point 
energy and luminosity errors which matter (about 1.3~MeV). The error on \S0 is 
dominated by the theoretical uncertainty on the small angle bhabha calculations 
(0.11 \%), but this is going to improve very soon with the new estimation 
of this uncertainty (0.06 \%) shown at this workshop~\cite{WARD}. Moreover, a QED 
uncertainty estimated to be around 0.05 \% has also been included in the fits.

The results of the lineshape fit are shown in table~\ref{tab:lineshape} with and 
without the hypothesis of lepton universality. From them, the leptonic widths 
and the invisible \Z0~width are derived.
 
\begin{table}[t]
\caption{Average line shape parameters from the results of the four LEP experiments.
\label{tab:lineshape}}
\vspace{0.2cm}
\begin{center}
\footnotesize
\begin{tabular}{|l|c|c|}
\hline
\raisebox{0pt}[13pt][7pt]{Parameter} &
\raisebox{0pt}[13pt][7pt]{Fitted Value} & 
\raisebox{0pt}[13pt][7pt]{Derived Parameters} \\
\hline
\MZ                       &    91186.7 $\pm$ 2.1~MeV  &  \\
\GZ                       &    2493.9  $\pm$ 2.4~MeV  &  \\
$\sigma^0_{\mathrm{had}}$ &    41.491  $\pm$ 0.058~nb &  \\
$\mathrm{R}_{\mathrm{e}}$ &    20.783  $\pm$ 0.052    & $\Gamma_{\mathrm{e}}$  = 83.87  $\pm$ 0.14~MeV \\
$\mathrm{R}_{\mu}$        &    20.789  $\pm$ 0.034    & $\Gamma_{\mu}$         = 83.84  $\pm$ 0.18~MeV \\
$\mathrm{R}_{\tau}$       &    20.764  $\pm$ 0.045    & $\Gamma_{\tau}$        = 83.94  $\pm$ 0.22~MeV \\
\hline
\multicolumn{3}{|c|}{\raisebox{0pt}[12pt][6pt]{With Lepton Universality}} \\
\hline
                          &                           & $\Gamma_{\mathrm{had}}$= 1742.3 $\pm$  2.3~MeV \\
$\mathrm{R}_{\mathrm{l}}$ &    20.765  $\pm$ 0.026    & $\Gamma_{\mathrm{l}}$  = 83.90  $\pm$ 0.10~MeV \\
                          &                           & $\Gamma_{\mathrm{inv}}$= 500.1  $\pm$  1.9~MeV \\
\hline
\end{tabular}
\end{center}
\end{table}

From the measurement of the \Z0~invisible width, and assuming the ratio of the partial widths to 
neutrinos and leptons to be the MSM predictions 
($\frac{\Gamma_{\nu}}{\Gamma_{\mathrm{l}}} = 1.991 \pm 0.001 $), the number of light neutrinos 
species is measured to be

\noindent
\bea
\mathrm{N}_{\nu} & = & 2.994 \pm 0.011. \nonumber
\eea

Alternatively, one can assume three neutrino species and determine the width from additional 
invisible decays of the \Z0~to be $\Delta\Gamma_{\mathrm{inv}}<2.8$~MeV~@95\%~C.L.

The measurement of \Rl~is very sensitive to PQCD corrections, thus it can be used to determine 
the value of \als. A combined fit to the measurements shown in table~\ref{tab:lineshape}, and 
imposing \mt=173.8$\pm$5.0~GeV as a constraint gives: 

\noindent
\bea
\alpha_{\mathrm{s}}(\mathrm{M}^2_{\mathrm{Z}}) & = & 0.123 \pm 0.004  \nonumber
\eea

\noindent in agreement with the world average~\cite{ALPHAS} 
$\alpha_{\mathrm{s}}(\mathrm{M}^2_{\mathrm{Z}}) = 0.119 \pm 0.002$. 

\subsection{\bf Heavy flavour results}\label{subsec:HF}

The large mass and long lifetime of the $b$ and $c$ quarks provides a way to perform flavour tagging. 
This allows for precise measurements of the partial widths of the decays \Z0$\rightarrow c \bar{c}$ and 
\Z0$\rightarrow b \bar{b}$. It is useful to normalise the partial width to \Gh~by 
measuring the partial decay fractions with respect to all hadronic decays

\noindent
\bea
\mathrm{R}_{\mathrm{c}} \equiv \frac{\Gamma_{c}}{\Gamma_{\mathrm{h}}} & , & 
\mathrm{R}_{\mathrm{b}} \equiv \frac{\Gamma_{b}}{\Gamma_{\mathrm{h}}}.  \nonumber 
\eea

With this definition most of the radiative corrections appear both in the numerator 
and denominator and thus cancel out, with the important exception of the vertex corrections 
in the \Z0 $b\bar{b}$ vertex. This is the only relevant correction to \Rb, and within the 
MSM basically depends on a single parameter, the mass of the top quark.

The partial decay fractions of the \Z0~to other quark flavours, like \Rc, are only weakly 
dependent on \mt; the residual weak dependence is indeed due to the presence of 
$\Gamma_{b}$ in the denominator. The MSM predicts \Rc = 0.172, valid over a wide 
range of the input parameters.

The combined values from the measurements of LEP and SLD gives

\noindent
\bea
\mathrm{R}_{\mathrm{b}} & = & 0.21656 \pm 0.00074 \nonumber \\
\mathrm{R}_{\mathrm{c}} & = & 0.1735  \pm 0.0044  \nonumber 
\eea

\noindent with a correlation of -17\% between the two values. 
%The large sensitivity 
%of \Rb~to the top mass allows to determine indirectly its mass to be \mt=151$\pm$25~GeV, 
%in agreement with the direct measurement (\mt=173.8$\pm$5.0~GeV).

\section{\bf \Z0~asymmetries: $\boldmath{\sin^2\theta_{\mathrm{eff}}}$ }

Parity violation in the weak neutral current is caused by the difference of couplings 
of the \Z0~to right-handed and left-handed fermions. If we define $A_{\mathrm{f}}$ as

\noindent
\bea
A_{\mathrm{f}} & \equiv & 
\frac{2 \biggl( \frac{g^f_V}{g^f_A} \biggr) }{1 + \biggl( \frac{g^f_V}{g^f_A}\biggr)^2},
\eea

\noindent where $g^f_{V(A)}$ denotes the vector(axial-vector) coupling constants, one can write 
all the \Z0~asymmetries in terms of $A_{\mathrm{f}}$.

Each process $e^+ e^- \rightarrow \mathrm{Z}^{0} \rightarrow \mathrm{f}\bar{\mathrm{f}}$ 
can be characterised 
by the direction and the helicity of the emitted fermion (f). Calling forward the hemisphere 
into which the electron beam is pointing, the events can be subdivided into four categories: 
FR,BR,FL and BL, corresponding to right-handed (R) or left-handed (L) fermions emitted 
in the forward (F) or backward (B) direction. Then, one can write three \Z0~asymmetries 
as:

\noindent
\bea
A_{\mathrm{pol}} \equiv & 
\frac{\sigma_{\mathrm{FR}}+\sigma_{\mathrm{BR}}-\sigma_{\mathrm{FL}}-\sigma_{\mathrm{BL}}}
     {\sigma_{\mathrm{FR}}+\sigma_{\mathrm{BR}}+\sigma_{\mathrm{FL}}+\sigma_{\mathrm{BL}}}
                        & = -A_{\mathrm{f}} \\
A^{\mathrm{FB}}_{\mathrm{pol}} \equiv & 
\frac{\sigma_{\mathrm{FR}}+\sigma_{\mathrm{BL}}-\sigma_{\mathrm{BR}}-\sigma_{\mathrm{FL}}}
     {\sigma_{\mathrm{FR}}+\sigma_{\mathrm{BR}}+\sigma_{\mathrm{FL}}+\sigma_{\mathrm{BL}}}
                        & = -\frac{3}{4} A_{\mathrm{e}}  \\
A_{\mathrm{FB}} \equiv & 
\frac{\sigma_{\mathrm{FR}}+\sigma_{\mathrm{FL}}-\sigma_{\mathrm{BR}}-\sigma_{\mathrm{BL}}}
     {\sigma_{\mathrm{FR}}+\sigma_{\mathrm{BR}}+\sigma_{\mathrm{FL}}+\sigma_{\mathrm{BL}}}
                        & = \frac{3}{4} A_{\mathrm{e}}A_{\mathrm{f}} 
\eea

\noindent and in case the initial state is polarised with some degree of polarisation 
($P$), one can define:

\noindent
\bea
A_{\mathrm{LR}} \equiv & \frac{1}{P} 
\frac{\sigma_{\mathrm{Fl}}+\sigma_{\mathrm{Bl}}-\sigma_{\mathrm{Fr}}-\sigma_{\mathrm{Br}}}
     {\sigma_{\mathrm{Fr}}+\sigma_{\mathrm{Br}}+\sigma_{\mathrm{Fl}}+\sigma_{\mathrm{Bl}}}
                        & = A_{\mathrm{e}} \\
A^{\mathrm{pol}}_{\mathrm{FB}} \equiv & -\frac{1}{P} 
\frac{\sigma_{\mathrm{Fr}}+\sigma_{\mathrm{Bl}}-\sigma_{\mathrm{Fl}}-\sigma_{\mathrm{Br}}}
     {\sigma_{\mathrm{Fr}}+\sigma_{\mathrm{Br}}+\sigma_{\mathrm{Fl}}+\sigma_{\mathrm{Bl}}}
                        & = \frac{3}{4} A_{\mathrm{f}} 
\eea

\noindent where r(l) denotes the right(left)-handed initial state polarisation. Assuming lepton 
universality, all this observables depend only on the ratio between the vector and axial-vector 
couplings. It is conventional to define the effective mixing angle \SEF~as

\noindent
\bea
\sin^2\theta_{\mathrm{eff}} & \equiv & \frac{1}{4} \biggl( 1 - \frac{g^l_V}{g^l_A}  \biggr)
\eea

\noindent and to collapse all the asymmetries into a single parameter \SEF.

\subsection{\bf Lepton asymmetries}\label{subsec:lepton_asym}
\subsubsection{\bf Angular distribution}\label{subsec:lepton_afb}

The lepton forward-backward asymmetry is measured from the angular distribution 
of the final state lepton.
The measurement of $A^{\mathrm{l}}_{FB}$ is quite simple and robust and its 
accuracy is limited by the statistical error. The common systematic uncertainty in the LEP measurement 
due to the uncertainty on the LEP energy measurement is about~0.0007.
The values measured by the LEP collaborations are in agreement with lepton universality,

\noindent
\bea
 & A^e_{\mathrm{FB}}     = 0.0153 \pm 0.0025  &    \nonumber \\
 & A^{\mu}_{\mathrm{FB}} = 0.0164 \pm 0.0013  &    \nonumber \\
 & A^{\tau}_{\mathrm{FB}}= 0.0183 \pm 0.0017  &    \nonumber 
\eea

\noindent and can be combined into a single measurement of \SEF,

\noindent
\bea
A^{\mathrm{l}}_{\mathrm{FB}}= 0.01683 \pm 0.00096 & \Longrightarrow &
{ \sin^2\theta_{\mathrm{eff}} = 0.23117 \pm 0.00054}. \nonumber 
\eea

\subsubsection{\bf Tau polarisation at LEP}\label{subsec:lepton_taupol}

Tau leptons decaying inside the apparatus acceptance can be used to measure 
the polarised asymmetries defined by equations~(4) and~(5). A more sensitive method 
is to fit the measured dependence of $A_{\mathrm{pol}}$ as a function of the 
polar angle $\theta$ :

\noindent
\bea
A_{\mathrm{pol}}(\cos\theta) & =  & 
- \frac{A_{\tau}(1+\cos^2\theta)+2 A_e \cos\theta}{(1+\cos^2\theta) + 2 A_{\tau} A_e \cos\theta }
\eea

The sensitivity of this measurement to \SEF~is larger because the dependence 
on $A_{\mathrm{l}}$ is linear to a good approximation. 
The accuracy of the measurements is dominated by the statistical 
error. The typical systematic error is about 0.003 for $A_{\tau}$ and 0.001 for $A_e$.
The LEP measurements are:

\noindent
\bea
A_{e}   = 0.1479 \pm 0.0051 & \Longrightarrow &
{ \sin^2\theta_{\mathrm{eff}} = 0.23141 \pm 0.00065} \nonumber  \\
A_{\tau}= 0.1431 \pm 0.0045 & \Longrightarrow &
{ \sin^2\theta_{\mathrm{eff}} = 0.23202 \pm 0.00057} \nonumber  
\eea

\subsection{\bf \ALR~from SLD}\label{subsec:alr}

The linear accelerator at SLAC (SLC) allows to collide positrons 
with a highly longitudinally polarised electron beam (up to 77\% polarisation). Therefore, 
the SLD detector can measure the left-right cross-section asymmetry (\ALR) 
defined by equation~(7). This observable is a factor of 4.6 times more sensitive to 
\SEF~than, for instance, $A^{\mathrm{l}}_{\mathrm{FB}}$ for a given precision.
The measurement is potentially free of experimental 
systematic errors, with the exception of the polarisation measurement that 
has been carefully cross-checked at the 1\% level. The last update on this 
measurement gives

\noindent
\bea
A_{\mathrm{LR}}   = 0.1510 \pm 0.0025 & \Longrightarrow &
{ \sin^2\theta_{\mathrm{eff}} = 0.23101 \pm 0.00031}, \nonumber 
\eea

\noindent and assuming lepton universality it can be 
combined with preliminary measurements at SLD of the lepton
left-right forward-backward asymmetry ($A^{\mathrm{pol}}_{\mathrm{FB}}$) 
defined in equation~(8) to give

\noindent
\bea
         & {\sin^2\theta_{\mathrm{eff}} = 0.23109 \pm 0.00029}. &  \nonumber 
\eea

\subsection{\bf Lepton couplings}\label{subsec:lepton_coup}

All the previous measurements of the lepton coupling ($A_{\mathrm{l}}$) can be 
combined with a $\chi^2/\mathrm{dof}=2.2/3$ and give
\noindent
\bea
A_{\mathrm{l}}   = 0.1489 \pm 0.0017 & \Longrightarrow &
{\sin^2\theta_{\mathrm{eff}} = 0.23128 \pm 0.00022}. \nonumber 
\eea
 
\begin{figure}
\centering
\mbox{%
\epsfig{file=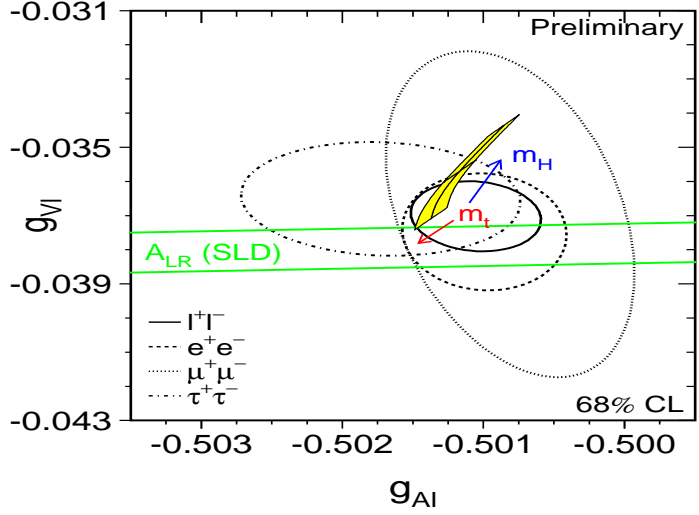 % this is the name of your figure file
        ,height=8cm  % this is the height of the figure (optional)
        ,width=10cm   % this is the width of the figure (optional)
       }%
}
\caption{Contours of 68\% probability in the $g^l_V - g^l_A$ plane. The solid 
contour results from a fit assuming lepton universality. Also shown is the one 
standard deviation band resulting from the \ALR~measurement of SLD.} % the figure caption
\label{fig:gvga}      % the (optional) label to refer to in the text
\end{figure}

The asymmetries measured are only sensitive to the ratio between the 
vector and axial-vector couplings. If we introduce also the measurement 
of the leptonic width shown in table~\ref{tab:lineshape} we can 
fit the lepton couplings to the \Z0~to be

\noindent
\bea
g^l_V & = & -0.03753 \pm 0.00044, \nonumber \\
g^l_A & = & -0.50102 \pm 0.00030, \nonumber 
\eea

\noindent where the sign is chosen to be negative by definition. 
Figure~\ref{fig:gvga} shows the 68~\% probability contours in the 
$g^l_V - g^l_A$ plane. 

%Notice that the fitted value of $g^l_A$ is 
%different from the Born prediction ($-\frac{1}{2}$) by about $3.4 \sigma$, 
%showing evidence for radiative corrections in the $\rho$ parameter.

\subsection{\bf Quark asymmetries}\label{subsec:quark_asym}

\subsubsection{\bf Heavy Flavour asymmetries}\label{subsec:HF_asym}

The inclusive measurement of the $b$ and $c$ asymmetries is more sensitive 
to \SEF~than, for instance, the leptonic forward-backward asymmetry. The 
reason is that $A_b$ and $A_c$ are mostly independent of \SEF, therefore 
$A^{b(c)}_{\mathrm{FB}}$ (which is proportional to the product $A_e A_{b(c)}$)
is a factor 3.3(2.4) more sensitive than $A^{\mathrm{l}}_{\mathrm{FB}}$.
The typical systematic uncertainty in $A^{b(c)}_{\mathrm{FB}}$ is 
about 0.001(0.002) and the precision of the measurement is  
dominated by statistics. 

SLD can measure also the $b$ and $c$
left-right forward-backward asymmetry defined in equation~(8) which is 
a direct measurement of the quark coupling $A_b$ and $A_c$. 
The combined fit for the LEP and SLD measurements gives

\noindent
\bea
 A^b_{\mathrm{FB}}= 0.0990 \pm 0.0021  & \Longrightarrow & {\sin^2\theta_{\mathrm{eff}} = 0.23225 \pm 0.00038} 
\nonumber \\
 A^c_{\mathrm{FB}}= 0.0709 \pm 0.0044  & \Longrightarrow & {\sin^2\theta_{\mathrm{eff}} = 0.2322 \pm 0.0010} 
\nonumber \\
 A_b              = 0.867  \pm 0.035   &   &  \nonumber \\
 A_c              = 0.647  \pm 0.040   &   &  \nonumber 
\eea

\noindent where 13\% is the largest correlation between $A^b_{\mathrm{FB}}$ and $A^c_{\mathrm{FB}}$.

Taking the value of $A_{\mathrm{l}}$ given in section~\ref{subsec:lepton_coup} and these 
measurements together in a combined fit gives

\noindent
\bea
 & A_b              = 0.881  \pm 0.018   &    \nonumber \\
 & A_c              = 0.641  \pm 0.028   &    \nonumber 
\eea

\noindent to be compared with the MSM predictions $A_b = 0.935$ and 
$A_c = 0.668$ valid over a wide 
range of the input parameters.
%where the uncertainty in the predictions due to the input parameters is negligible. 
The measurement of $A_c$ is in good agreement with expectations,
while the measurement of $A_b$ is 3 standard deviations lower than 
the predicted value. 
This is due to three independent measurements: the SLD measurement of 
$A_b$ is low compared with the MSM, while the LEP measurement of 
$A^b_{\mathrm{FB}}$ is low and the SLD measurement of 
$A_{\mathrm{LR}}$ is high compared with the results of the best fit to 
the MSM predictions (see section~\ref{subsec:fit}).

%This is due to three independent measurements: the 
%LEP measurement of $A^b_{\mathrm{FB}}$ is low compared with the MSM, 
%the SLD measurement of $A_b$ is also low, and the SLD measurement of 
%$A_{\mathrm{LR}}$ is high compared with the MSM.

\subsubsection{\bf Jet charge asymmetries}\label{subsec:jet_asym}

The average charge flow in the inclusive samples of hadronic \Z0~decays 
is related to the forward-backward asymmetries of individual quarks:

\noindent
\bea
 \langle \mathrm{Q}_{\mathrm{FB}} \rangle & = & \sum_{\mathrm{q}} 
\delta_{\mathrm{q}} A^{\mathrm{q}}_{\mathrm{FB}} \frac{\Gamma_{\mathrm{q\bar{q}}}}{\Gamma_{\mathrm{h}}}   
\eea

\noindent where $\delta_{\mathrm{q}}$, the charge separation, is the average charge 
difference between the quark and antiquark hemispheres in an event.
The combined LEP value is

\noindent
\bea
         & {\sin^2\theta_{\mathrm{eff}} = 0.2321 \pm 0.0010}. &  \nonumber 
\eea

\subsubsection{\bf Comparison of the determinations of 
$\boldmath{\sin^2\theta_{\mathrm{eff}}}$}\label{subsec:sineff}

The combination of all the quark asymmetries shown in this section can be 
directly compared to the determination of \SEF~obtained with leptons, 

\noindent
\bea
{\sin^2\theta_{\mathrm{eff}} = 0.23222 \pm 0.00033} &   & \mathrm{(quark-asymmetries)} \nonumber \\ 
{\sin^2\theta_{\mathrm{eff}} = 0.23128 \pm 0.00022} &   & \mathrm{(lepton-asymmetries)} \nonumber 
\eea

\noindent which shows a 2.4~$\sigma$ discrepancy.

\begin{figure}
\centering
\mbox{%
\epsfig{file=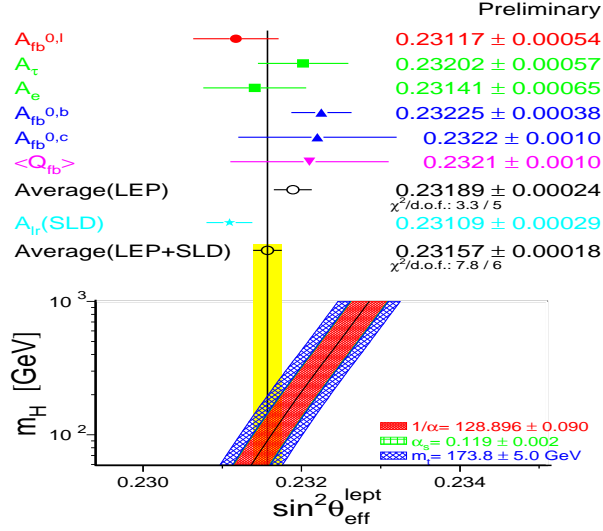 % this is the name of your figure file
        ,height=7cm  % this is the height of the figure (optional)
        ,width=8cm   % this is the width of the figure (optional)
       }%
}
\caption{Comparison of several determinations of \SEF~from asymmetries.} % the figure caption
\label{fig:sef}      % the (optional) label to refer to in the text
\end{figure}

Over all, the agreement is very good, and the combination of the individual 
determinations of \SEF~gives

\noindent
\bea
         & {\sin^2\theta_{\mathrm{eff}} = 0.23157 \pm 0.00018} &  \nonumber 
\eea

\noindent with a $\chi^2/\mathrm{dof}=7.8/6$ as it is shown 
in figure~\ref{fig:sef}.

%\section{\MW and \SW}
\section{Consistency with the Minimal Standard Model}\label{sec:consistency}

The MSM predictions are computed using the programs TOPAZ0~\cite{TOPAZ0} 
and ZFITTER~\cite{ZFITTER}. They represent the state-of-the-art in the 
computation of radiative corrections, and incorporate recent calculations such as
the QED radiator function to \cal{O}($\alpha^3$)~\cite{ALPHA3}, four-loop 
QCD effects~\cite{QCD4LOOP}, non-factorisable QCD-EW corrections~\cite{QCDEW}, 
and two-loop sub-leading 
\cal{O}($\alpha^2 \mathrm{m}^2_{\mathrm{t}} / \mathrm{M}^2_{\mathrm{Z}}$) 
corrections~\cite{DEGRASSI}, resulting in a significantly reduced theoretical 
uncertainty compared to the work summarized in reference~\cite{YR95}.

\subsection{\bf Are we sensitive to radiative corrections other than $\Delta\alpha$?}\label{subsec:sensitivity}

This is the most natural question to ask if one pretends to test the MSM as 
a Quantum Field Theory and
to extract information on the only unknown parameter in the MSM, \MH.
%In fact, the answer is already given in table~\ref{tab:sensitiv}, 
%where one can see the different sensitivity of the MSM prediction on the 
%input parameters. 

The MSM prediction of \Rb~neglecting radiative corrections is ${\mathrm{R}}^0_{\mathrm{b}}=0.2183$, 
while the measured value given in section~\ref{subsec:HF} is about $2.3\sigma$ lower.
From table~\ref{tab:sensitiv} one can see that the MSM prediction depends only 
on \mt~and allows to determine indirectly its mass to be \mt=151$\pm$25~GeV, 
in agreement with the direct measurement (\mt=173.8$\pm$5.0~GeV).

%The large sensitivity 
%of \Rb~to the top mass allows to determine indirectly its mass to be \mt=151$\pm$25~GeV, 
%in agreement with the direct measurement (\mt=173.8$\pm$5.0~GeV).

From the measurement of the leptonic width, the vector-axial coupling given in 
section~\ref{subsec:lepton_coup} disagrees with the Born prediction (-1/2) by about 
$3.4\sigma$, showing evidence for radiative corrections in the 
$\rho$ parameter, $\Delta\rho = 0.0041 \pm 0.0012$.

%Notice that the fitted value of $g^l_A$ is 
%different from the Born prediction ($-\frac{1}{2}$) by about $3.4 \sigma$, 
%showing evidence for radiative corrections in the $\rho$ parameter.

However, the most striking evidence for pure weak radiative corrections is not coming 
from \Z0~physics, but from \MW~and its relation with \GM. The value measured 
at LEP and TEVATRON~\cite{LANCON} is \MW=$80.39 \pm 0.06$~GeV. From this measurement 
and through the relation

\noindent
\bea
 \mathrm{M}^2_{\mathrm{W}} \left(1 - \frac{\mathrm{M}^2_{\mathrm{W}}}{\mathrm{M}^2_{\mathrm{Z}}} \right) & 
 = & \frac{\pi \alpha}{G_{\mu} \sqrt{2}} \left( 1 + \Delta r\right)  
\eea

\noindent one gets a measurement of $\Delta r = 0.036 \pm 0.004$, and subtracting the 
value of $\Delta\alpha$ ($\Delta\alpha = -\Pi_{\gamma\gamma}$),
given in section~\ref{subsec:input_par}, 
one obtains $\Delta r_{\mathrm{W}} = \Delta r - \Delta \alpha = -0.027 \pm 0.004$, 
which is about $6.8\sigma$ different from zero. A more detailed investigation 
on the evidence for pure weak radiative corrections can be found in reference~\cite{SIRLIN}.

\subsection{\bf Fit to the MSM predictions}\label{subsec:fit}

\begin{figure}
\centering
\mbox{%
\epsfig{file=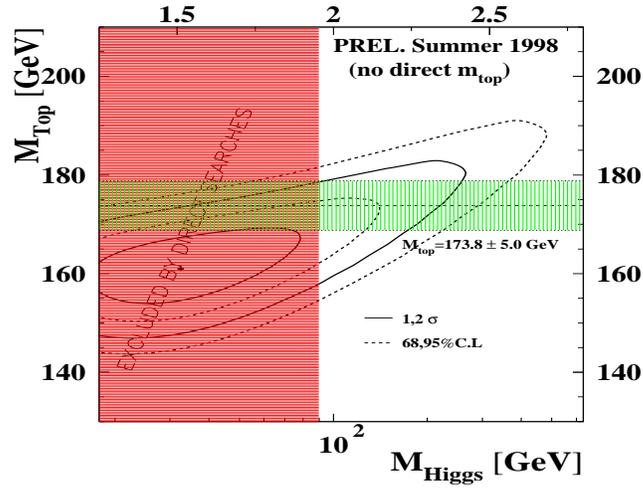 % this is the name of your figure file
        ,height=7cm  % this is the height of the figure (optional)
        ,width=9cm   % this is the width of the figure (optional)
       }%
}
\caption{The 68\% and 95\% confidence level contours in the \mt~vs~\MH~plane. The 
vertical band shows the 95\% C.L. exclusion limit on \MH~from direct searches.} % the figure caption
\label{fig:mhmt}      % the (optional) label to refer to in the text
\end{figure}

\begin{figure}
\centering
\mbox{%
\epsfig{file=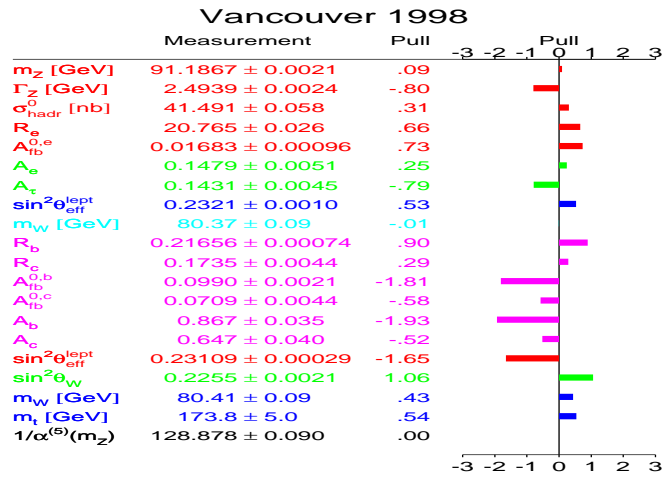 % this is the name of your figure file
        ,height=7cm  % this is the height of the figure (optional)
        ,width=9cm   % this is the width of the figure (optional)
       }%
}
\caption{Pulls of the measurements with respect to the best fit results. The 
pull is defined as the difference of the measurement to the fit prediction divided 
by the measurement error.} % the figure caption
\label{fig:pulls}      % the (optional) label to refer to in the text
\end{figure}

Having shown that there is sensitivity to pure weak corrections with the accuracy in the 
measurements achieved so far, one can envisage to fit the values of the unknown Higgs mass and 
the less well known top mass in the context of the MSM predictions. The fit 
is done using not only the \Z0~measurements shown in this letter but also using 
the W mass measurements~\cite{LANCON} and $\nu$N scattering measurements~\cite{NUTEV}.
The quality of the fit is very good, ($\chi^2/\mathrm{dof}=13.3/14$) and 
the result is,

\noindent
\bea
\mathrm{m}_{\mathrm{t}}      & = & 161.1^{+8.2}_{-7.1}~\mathrm{GeV}                               \nonumber 
%\mathrm{m}_{\mathrm{t}}       = & 161.1^{+8.2}_{-7.1} \mathrm{GeV} &                              \nonumber \\ 
%\log(\mathrm{M}_{\mathrm{H}}) = & 1.51^{+0.38}_{-0.29}             & (32^{+46}_{-16} \mathrm{GeV}) \nonumber \\ 
%\alpha_s                      = & 0.120 \pm 0.003                  &  \nonumber
\eea

\noindent to be compared with \mt=173.8$\pm$5.0~GeV measured at TEVATRON. The result 
of the fit is shown in the \MH-\mt~plane in figure~\ref{fig:mhmt}. Both 
determinations of \mt~have similar precision and are compatible ($1.4\sigma$). Therefore, 
one can constrain the previous fit with the direct measurement of \mt~and obtains:

\noindent
\bea
\mathrm{m}_{\mathrm{t}}       = & 171.1 \pm 4.9~\mathrm{GeV} &                              \nonumber \\ 
\log(\mathrm{M}_{\mathrm{H}}/\mathrm{GeV}) = & 1.88^{+0.33}_{-0.41}       & 
                                   (\mathrm{M}_{\mathrm{H}} = 76^{+85}_{-47}~\mathrm{GeV}) \nonumber \\ 
\alpha_s                      = & 0.119 \pm 0.003            &  \nonumber
\eea

\noindent with a very reasonable $\chi^2/\mathrm{dof}=14.9/15$. The agreement 
of the fit with the measurements is impressive and it is shown as a pull distribution in 
figure~\ref{fig:pulls}. 

The best indirect determination of the W mass is obtained from 
the MSM fit when no information from the direct measurement is used, 

\noindent
\bea
\mathrm{M}_{\mathrm{W}}     &  =  & 80.367 \pm 0.029~\mathrm{GeV}. \nonumber
\eea
%Therefore 
%the aim for LEP2 and TEVATRON is to measure the W mass with at least a precision 
%of 30~MeV.

The most significant correlation on the fitted parameters is 77\% between 
$\log(\mathrm{M}_{\mathrm{H}}/\mathrm{GeV})$ and $\alpha(\mathrm{M}^2_{\mathrm{Z}})$. 
If one of the more precise new evaluations of $\Delta\alpha$ mentioned in 
section~\ref{subsec:input_par} is used, this correlation decreases dramatically 
and the precision on $\log(\mathrm{M}_{\mathrm{H}}/\mathrm{GeV})$ improves by about 30\%. 
For instance, using $\alpha^{-1}(\mathrm{M}^2_{\mathrm{Z}}) = 128.923 \pm 0.036$ 
from reference~\cite{DAVIER}, one gets: 

\noindent
\bea
\mathrm{m}_{\mathrm{t}}       = & 171.4 \pm 4.8~\mathrm{GeV}       &                              \nonumber \\ 
\log(\mathrm{M}_{\mathrm{H}}/\mathrm{GeV}) = & 1.96^{+0.23}_{-0.26}       & 
                                   (\mathrm{M}_{\mathrm{H}} = 91^{+64}_{-41}~\mathrm{GeV}) \nonumber \\ 
\alpha_s                      = & 0.119 \pm 0.003                  &  \nonumber
\eea

\noindent with the same confidence level ($\chi^2/\mathrm{dof}=14.9/15$) and 
a correlation of 39\% between 
$\log(\mathrm{M}_{\mathrm{H}}/\mathrm{GeV})$ and $\alpha(\mathrm{M}^2_{\mathrm{Z}})$.

\section{Constraints on \MH}

In the previous section it has been shown that the global MSM fit to the data
gives
\noindent
\bea
\log(\mathrm{M}_{\mathrm{H}}/\mathrm{GeV}) = & 1.88^{+0.33}_{-0.41}       & 
                                   (\mathrm{M}_{\mathrm{H}} = 76^{+85}_{-47}~\mathrm{GeV}) \nonumber 
\eea

\noindent and taking into account the theoretical uncertainties (about $\pm$0.05 in 
$\log(\mathrm{M}_{\mathrm{H}}/\mathrm{GeV})$), this implies a one-sided 95\%~C.L. limit of:

\noindent
\bea
\mathrm{M}_{\mathrm{H}} &  < & 262~\mathrm{GeV}~@ 95 \%~\mathrm{C.L.}  \nonumber 
\eea

\noindent which does not take into account the limits from direct searches 
($\mathrm{M}_{\mathrm{H}} > 89.8~\mathrm{GeV}~@ 95 \%~\mathrm{C.L.}$).

\subsection{\bf Is this low value of \MH~a consequence of a particular measurement?}\label{subsec:consistency}

As described in section~\ref{subsec:what_meas}, one can divide the measurements sensitive 
to the Higgs mass into three different groups: Asymmetries, Widths and the W mass. They test 
conceptually different components of the radiative corrections and it is interesting to 
check the internal consistency.

\begin{figure}
\centering
\mbox{%
\epsfig{file=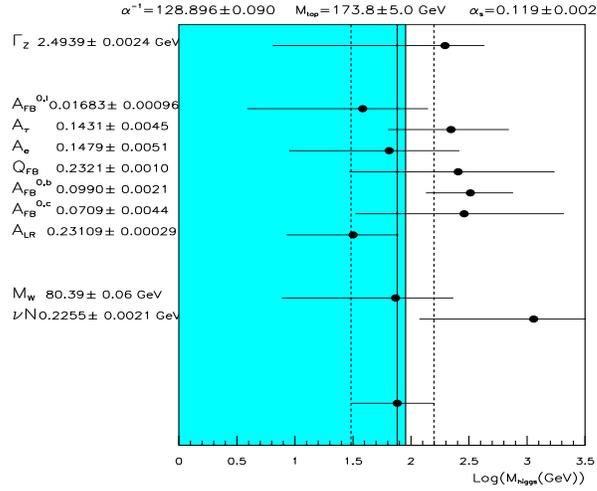 % this is the name of your figure file
        ,height=7cm  % this is the height of the figure (optional)
        ,width=9cm   % this is the width of the figure (optional)
       }%
}
\caption{Individual determination of $\log(\mathrm{M}_{\mathrm{H}}/\mathrm{GeV})$ for each 
of the measurements.} % the figure caption
\label{fig:logmh}      % the (optional) label to refer to in the text
\end{figure}

Repeating the MSM fit shown in the previous section for the three different groups of measurements
with the additional constraint from~\cite{ALPHAS} $\alpha_s = 0.119 \pm 0.002$ gives the results shown in
the second column in 
table~\ref{tab:consistency}. All the fits are consistent with a low value of the Higgs mass, 
and there is no particular set of measurements that pulls \MH~down. This is seen with even more 
detail in figure~\ref{fig:logmh}, where the individual determinations of 
$\log(\mathrm{M}_{\mathrm{H}}/\mathrm{GeV})$ are shown for each measurement.

\begin{table}[t]
\caption{ Results on $\log(\mathrm{M}_{\mathrm{H}}/\mathrm{GeV})$ for different samples of 
measurements. In the fit the input parameters and their uncertainties are taken to 
be the values presented in section~\ref{subsec:input_par}. The impact of the uncertainty in each 
parameter is explicitely shown. \label{tab:consistency}}
\vspace{0.2cm}
\begin{center}
\footnotesize
\begin{tabular}{|l|c|c c c c c c c c c|}
\hline
{} &\raisebox{0pt}[13pt][7pt]{$\log(\mathrm{M}_{\mathrm{H}}/\mathrm{GeV})$} &
\raisebox{0pt}[13pt][7pt]{$[\Delta\log(\mathrm{M}_{\mathrm{H}}/\mathrm{GeV})]^2$}    & = &
\raisebox{0pt}[13pt][7pt]{$[\Delta_{\mathrm{exp.}}]^2 $}                & + &
\raisebox{0pt}[13pt][7pt]{$[\Delta\mathrm{m}_{\mathrm{t}}]^2 $}         & + &
\raisebox{0pt}[13pt][7pt]{$[\Delta\alpha]^2 $}                          & + &
\raisebox{0pt}[13pt][7pt]{$[\Delta\alpha_s]^2$} \\
\hline
\Z0~Asymmetries                                    & $1.94^{+0.29}_{-0.31}$ & 
$[0.29]^2$ & = & $[0.19]^2$ & + & $[0.12]^2$ & + & $[0.19]^2$ & + & $[0.01]^2$     \\ 
\Z0~Widths                                         & $2.21^{+0.36}_{-1.43}$ &  
$[0.36]^2$ & = & $[0.31]^2$ & + & $[0.14]^2$ & + & $[0.08]^2$ & + & $[0.13]^2$     \\ 
$\mathrm{M}_{\mathrm{W}}$~and~$\nu\mathrm{N}$      & $2.04^{+0.45}_{-0.84}$ &
$[0.45]^2$ & = & $[0.41]^2$ & + & $[0.18]^2$ & + & $[0.08]^2$ & + & $[0.00]^2$     \\ 
\hline
\end{tabular}
\end{center}
\end{table}

\begin{figure}
\centering
\mbox{%
\epsfig{file=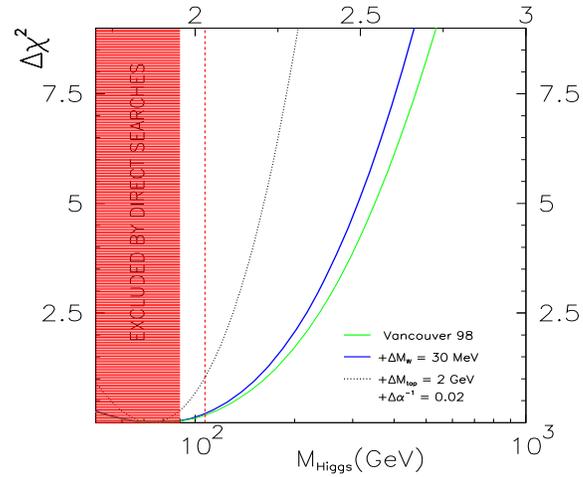 % this is the name of your figure file
        ,height=7cm  % this is the height of the figure (optional)
        ,width=8cm   % this is the width of the figure (optional)
       }%
}
\caption{ $\Delta\chi^2 = \chi^2 - \chi^2_{\mathrm{min}}$ vs. \MH~curve. Different 
cases are considered: the present situation, the future situation when 
$\Delta$\MW~is measured with a precision of 30~MeV and 
when $\Delta\alpha^{-1}=0.02$ and $\Delta$\mt=2~GeV.
The band shows the limit from direct searches, and the discontinous line the 
expected limit at the end of LEP2.} % the figure caption
\label{fig:chi2mh}      % the (optional) label to refer to in the text
\end{figure}

\subsection{\bf Is there any chance to improve these constraints?}\label{subsec:future}

%In table~\ref{tab:consistency} one can see also the different contributions to 
%$[\Delta\log(\mathrm{M}_{\mathrm{H}})]$. 

Although the most precise determination 
of the Higgs mass is still coming from the \Z0~asymmetries, it is clear 
from table~\ref{tab:consistency} that any 
future improvement will be limited by the uncertainty in 
$\alpha(\mathrm{M}^2_{\mathrm{Z}})$. If 
$\Delta\alpha^{-1} \sim 0.02$, then the error on 
$\log(\mathrm{M}_{\mathrm{H}}/\mathrm{GeV})$ is reduced to about 0.23 
($[\Delta_{\mathrm{exp.}}] \oplus [\Delta\mathrm{m}_{\mathrm{t}}]$), coming 
only from the \Z0~asymmetries measurements.

The accuracy of the W-boson mass is going to improve by a significant
factor in the near future. However, even if the W mass is measured with a 
precision of 30~MeV, the error on $\log(\mathrm{M}_{\mathrm{H}}/\mathrm{GeV})$ 
is going to be dominated by $\Delta\mathrm{m}_{\mathrm{t}}$ and will 
not be better than 0.23 obtained with the \Z0~asymmetries only,  
both determinations being highly correlated by the uncertainty on 
the top mass. 

Therefore, the error on $\log(\mathrm{M}_{\mathrm{H}}/\mathrm{GeV})$~is not going to 
improve significantly until a precise measurement of the top mass (2~GeV) 
becomes available. In such a case, one can easily obtain a precision 
in $\log(\mathrm{M}_{\mathrm{H}}/\mathrm{GeV})$ close to 0.15. This is what is shown 
in figure~\ref{fig:chi2mh}. 

Also shown in figure~\ref{fig:chi2mh} is the 
expected direct search limit from LEP2. If the tendency to prefer a very 
low value of \MH~continues with the new or updated measurements, and the accuracy  
on the top mass  and W-boson mass are improved significantly, consistent with the indirect 
determinations, we may be able to constrain severely the value of \MH.
%Then, either the Higgs is found at LEP2 (or TEVATRON) or we will have 
%an inconsistency of the MSM predictions. In any case, both situations look very 
%interesting.

% like the MSM predictions having difficulties 
%to describe our precision measurements!

\section{Conclusions and outlook}

The measurements performed at LEP and SLC have substantially improved 
the precision of the tests of the MSM, at the level of \cal{O}(0.1\%).
The effects of pure weak corrections are visible with a significance 
larger than three standard deviations from \Z0~observables and about 
seven standard deviations from the W-boson mass.

The top mass predicted by the MSM fits, (\mt=$161.1^{+8.2}_{-7.1}$~GeV) is
compatible (about $1.4\sigma$) with the direct measurement 
(\mt=$173.8 \pm 5.0$~GeV) and of similar precision.

The W-boson mass predicted by the MSM fits, ($\mathrm{M}_{\mathrm{W}}=80.367\pm0.029~\mathrm{GeV}$) 
is in very good agreement with the direct measurement 
($\mathrm{M}_{\mathrm{W}}=80.39\pm0.06~\mathrm{GeV}$).

The mass of the Higgs boson is predicted to be low,

\noindent
\bea
\log(\mathrm{M}_{\mathrm{H}}/\mathrm{GeV}) = & 1.88^{+0.33}_{-0.41}       & 
                                   (\mathrm{M}_{\mathrm{H}} = 76^{+85}_{-47}~\mathrm{GeV}) \nonumber \\ 
\mathrm{M}_{\mathrm{H}} &  < & 262~\mathrm{GeV}~@ 95 \%~\mathrm{C.L.}  \nonumber 
\eea

\noindent This uncertainty is reduced to $\Delta(\log(\mathrm{M}_{\mathrm{H}}/\mathrm{GeV})) \sim 0.23$ when
the uncertainty from 
$\Delta\alpha$ is negligible, and will be further reduced to  
$\Delta(\log(\mathrm{M}_{\mathrm{H}}/\mathrm{GeV})) \sim 0.15$ when \mt~is known with a 2~GeV precision 
and \MW~is known with a 30~MeV precision.

\section*{Acknowledgments}

I would like to thank Prof. Joan Sol\`{a} and all the organizing committee for the 
excellent organization of the workshop.
I'm very grateful to Martin W. Gr$\ddot{\mathrm{u}}$newald and Gunter Quast for his help 
in the preparation of the numbers and plots shown in this paper. I also thank
Guenther Dissertori for reading the paper and giving constructive criticisms.

%This is where one places acknowledgments for funding
%bodies etc.  Note that there are no section numbers for
%the Acknowledgments, Appendix or References.

%\section*{Appendix}

\end{document}